\documentstyle[11pt]{article}

\parskip=\medskipamount
\begin{document}
\rightline{\scriptsize{gr-qc/9605032}}
\begin{verbatim}

                       LA VERDAD OS HARA LIBROS

                             IFUG 94 - 12

\end{verbatim}

\vspace{1cm}
\begin{center}

{\large{\bf On the Estimates to Measure Hawking Effect and Unruh Effect
in the Laboratory}}\\
[.1in]
\vspace*{1.5cm}
{\bf H. C. Rosu \footnote{
internet: rosu@ifug.ugto.mx; IGSS, Magurele-Bucharest, Romania

PACS $\#$s : 04.20.Cv, 04.90.+e, 06.90.+v
}}

{\it Instituto de F\'{\i}sica de la Universidad de Guanajuato,\\
Apartado Postal E-143, 37150 Le\'on, Gto, M\'exico}\\
[-.03in]
\vspace*{2cm}
\vspace*{1cm}

{\it Honorable mention at 1992 Gravity Essays\\
                  slightly revised}\\
{\bf Abstract}

A comparison between the proposals made to measure Hawking-like effects
and the Unruh effect in the laboratory is given at the level of their
 estimates. No satisfactory scheme exists as yet for their detection.

\vspace*{1.5cm}

Int. J. Mod. Phys. D 3 (1994) 545-548

\end{center}
\vspace*{1cm}
\newpage
Two fundamental effects in present-day theoretical physics are
Hawking effect \cite{kn:h} and the very similar Unruh effect \cite{kn:u}.
They are
thermal-like effects involving microscopic (brownian) degrees of
freedom of quantum fields which are not causally connected.
 Thus the whole topic is
a fascinating and very challenging one for the physical interpretation.

Apart from the Hawking effect from primordial black holes, \cite{kn:-1},
 a number of experiments have been proposed in the
last decade to measure analogs of these effects in terrestrial
 laboratories even
though they are extremely small for any conceivable scheme. Indeed in
cgs units the relationship between the temperature parameter of
the effects and the proper acceleration parameter is:
$$ T=\frac{\hbar}{2\pi c k_{b}}a= 4\cdot 10^{-23} a       \eqno(1)$$
and therefore a thermal effect of only 1 K is produced by an
 acceleration of $2.4\cdot 10^{22} cm/s^{2}$. Clearly one is dealing
  with
effects which might have some experimental evidence only in very
 uncommon
situations like black-hole physics and/or ultrarelativistic non-linear
electrodynamics \cite{kn:md}. Nevertheless the analogies developed over
the years
showed that other fields of physics could have a contribution to
the better understanding of the two effects. Moreover, as a
corollary, those fields of physics enriched themselves with some
unconventional pictures.

With this in mind we pass to the scope of the note which is a critical
analysis
of the experimental estimates of the above mentioned fundamental
effects.

Unruh, \cite{kn:1}, was the first to propose an experimental
 scheme based on a
hydrodynamical analog of the Schwarzschild metric. He showed that
for a spherically symmetric, static convergent flow of an irrotational
fluid exceeding the speed of sound at some radius, the metric has
approximately the form of the Schwarzschild metric. He then quantized
the sound field and found an outgoing thermal flux of phonons emitted
by the sonic horizon at a temperature:
$$T=\hbar /2\pi k_{b}\cdot \partial v/\partial r = 10^{-2}K
(\frac{\partial v}{\partial r}/\frac{100m/s}{1\AA}).     \eqno(2)$$
The estimate (2) is very disappointing. To have 1 K one should produce
a velocity gradient of 100 m/s per $\AA$ at the sonic horizon. It is
by far doubtful that an atomic fluid could allow such huge gradients.
The situation may change in the case of superfluids,\cite{kn:0}.

In a series of papers written between 1983 and 1987,
Bell and Leinaas ,\cite{kn:2}, have proposed to interpret
the depolarization of electrons in storage rings
as a kind of circular Unruh effect. In the case of LEP the centripetal
acceleration is $a_{LEP}= 3\cdot 10^{22} g_{\oplus}$ implying a Unruh
temperature of 1200 K. This is already a measurable thermal-like
 effect.
However the depolarization of electons in storage rings has
a standard interpretation in terms of the well-known Sokolov-Ternov
 effect in
QED. Besides, the ``circular Unruh effect" is not as simple as the linear
one. Moreover, the thermal interpretation is questionable in circular motion,
\cite{kn:3}.
Bell and Leinaas have obtained a more complicated evolution of the
depolarization very close to isolated first order vertical betatron
resonances for perfectly aligned weak focusing storage rings.
 This situation is still to be
 checked, even though it is rather far from the experimental
  conditions at existing storage rings.

One of the best experimental scheme to measure the ``circular Unruh
 effect" belongs to Rogers, \cite{kn:4}. He proposed a small
  superconducting Penning
 trap with only one electron circulating inside, put into a microwave
 cavity. The ideal Penning trap is the mathematical problem of the
 electron motion in an external electromagnetic field consisting of a
 strong, uniform, axial magnetic field, and a quadrupole static electric
 field. Such a combination of fields is achieved in praxis by means
 of hyperbola-shaped electrodes. The equation of motion for the ideal
 fields is linear and could be expressed in terms of three normal
 modes , the axial, the magnetron, and the trap cyclotron modes.
 The usual working regime of the Penning trap is:
 $\omega _{tc} \gg \omega _{z} \gg \omega _{m}$, where the subscripts
 mean trap cyclotron, axial, and magnetron modes respectively .
 In the limit of vanishing electric field the trap cyclotron frequency
 goes into the free space cyclotron frequency, but otherwise it depends
 on the axial mode too.
 A transfer of the cyclotron vacuum noise energy to the lowest
  transverse magnetic mode of the
 microwave cavity is made via the axial resonant coupling.
  The acceleration in
 Rogers' experiment is $a= 6\cdot 10^{19} g_{\oplus}$ corresponding to
 T= 2.4 K. The critique of Rogers scheme
 is similar to that made above
 for storage rings. The circular noise is not universal and not very
  adaptable to a pure thermal interpretation, \cite{kn:3}. For further
  details on the Penning vacuum noise see \cite{kn:-3}.

Yablonovitch, \cite{kn:5}, proposed a plasma front generated
 on a subpicosecond time scale
as an accelerating fast-moving mirror. In this case vacuum fluctuations
near the plasma are subjected to the accelerating conversion into real
photons, and an Unruh effect of the non-adiabatic Casimir-type is
natural. Laser pulses of $10^{-12}$ sec could produce plasmas moving
with $a=10^{20}g_{\oplus}$ implying a Unruh temperature T= 4 K. The
plasma fronts could be generated by the non-adiabatic photoionization
of a gas or a semiconductor crystal.

Very interesting are the proposals of the YERPHY group. They showed that
channeling phenomena could be studied in the Unruh perspective too,
\cite{kn:6}.
The strong fields of crystalline axes and planes are acting with
 extremely large
transverse accelerations reaching $10^{33} cm/s^{2}$ in the
instantaneous rest frame of the ultrarelativistic
channeled particles ($\gamma =10^{8}$). The Armenian group provided
 an analysis of the Unruh radiation
of the channeled particles which arises as a result of the Compton
 scattering on the Planck spectrum of vacuum photons.
However only at $\gamma = 10^{8}$ the intensity per unit pathlength
of the Compton
 scattering on the Planck vacuum spectrum is comparable with the
 Bethe-Heitler brehmsstrahlung.

In another work, \cite{kn:7}, the YERPHY group discussed in
 the same spirit
the radiation of TeV electrons with initial velocity
perpendicular to a uniform magnetic field. The centripetal
acceleration is $a=\gamma e H/m\beta$ giving a Unruh temperature
$T=\gamma e H/2\pi k m \beta$.
This time the background which is the synchrotron radiation is surpassed
 by the Unruh radiation only at $\gamma = 10^{9}$ for a magnetic field
 $H=5\cdot 10^{7}G$, thus making impossible to detect Unruh radiation,
 say, at CLIC. Supercolliders with bunch structure capable of
 producing fields of the order $10^{9}G$ are required.

Another case considered in the same work is the propagation of an
electron through the electromagnetic field of a circularly
polarized plane wave. The proper centripetal acceleration is
 $a=2\omega \gamma\eta\sqrt{1+\eta^{2}}$, where $\omega$ is the
angular frequency of the electromagnetic wave,
 and $\eta = e\epsilon /m\omega$ ($\epsilon$ being the
amplitude of the field). The Unruh temperature is given by
$T=(\gamma \omega /\pi k)\eta \sqrt{1+\eta^{2}}$. The YERPHY group
calculated again the dependences of the intensity per unit
pathlength for Unruh radiation and for the radiation in the field
of an intense laser beam on x (the fraction of the initial energy taken
away by the radiated quanta) and on the parameter $\eta$.
Under such conditions the Unruh radiation might be observable already at
$\gamma \ge 10^{7}$.
Anyway, only a collider mode in combination with a powerful
 laser might make the Unruh signal available in such setups.

The last proposal is that involving the electromagnetic analog
of the Mach horizon, i.e., the Cherenkov effect in a GRIN (graded index)
dielectric material,\cite{kn:8}.
For such a scheme the estimated temperature is given by
$$T=\frac{\hbar c}{2\pi k_{b}}\frac{dn}{dr}.    \eqno(3)$$
The present-day optical gradients (0.2 mm$^{-1}$)
could generate a thermal effect of 0.7K. Again a rather disappointing
estimate. Besides, the so-called melting of the Cherenkov cone,
\cite{kn:9},that is
the structure of the distorted Cherenkov wavefront in GRIN materials
must be studied in great detail.

{\bf Acknowledgements}

The author would like to thank Professor Abdus Salam, the
 International Atomic
Energy agency and UNESCO for hospitality at the International
Centre for Theoretical Physics, Trieste, where the first version of
this work has been accomplished.

Partial support was provided by CONACyT Grants Nos.  1683-E9209 and
F246-E9207 to the University of Guanajuato.

\end{document}